\documentclass[a4paper,12pt]{article}
\usepackage{fullpage,cite}
\usepackage{graphicx}

\begin{document}

\title{Density-of-states picture and stability of ferromagnetism in the
highly-correlated Hubbard model}
\author{V.Yu. Irkhin and A.V. Zarubin$^{*}$ \\
Institute of Metal Physics, 620219 Ekaterinburg, Russia}
\maketitle

\begin{abstract}
The problem of stability of saturated and non-saturated
ferromagnetism in the Hubbard model is considered in terms of the
one-particle Green's functions. Approximations by Edwards and
Hertz and some versions of the self-consistent approximations
based on the $1/z$-expansion are considered. The account of
longitudinal fluctuations turns out to be essential for
description of the non-saturated state. The corresponding pictures
of density of states are obtained. ``Kondo'' density-of-states
singularities owing to spin-flip processes are analyzed. The
critical electron concentrations for instabilities of saturated
ferromagnetism and paramagnetic state are calculated for various
lattices. Drawbacks of various approximations are discussed. A
comparison with the results of previous works is performed.

71.10.Fd Lattice fermion models (Hubbard model, etc.), 71.28.+d
Narrow-band systems; intermediate-valence solids,
\end{abstract}

\section{Introduction}

The problem of ferromagnetic ordering in narrow energy bands is up
to now extensively discussed. Despite the large number of
publications on the topic, the magnetism of highly-correlated
electronic systems described by the Hubbard
model~\cite{Hubbard-I:1963} remains at the center of
attention~\cite{Izyumov, Shastry:1990, Linden:1991, Haaf:1992,
Haaf:1995, Strack:1995, Wurth:1996, Liang:1995, Obermeier:1997,
Vollhardt:1997}. Physically, in this case the picture of magnetism
is characterized by the existence of local magnetic moments and
differs substantially from the Stoner picture of a weak itinerant
magnetism~\cite{Moriya:1985}.

According to Nagaoka~\cite{Nagaoka:1966}, in the limit of infinite
Hubbard repulsion the ground state for simple lattices is a
saturated ferromagnetic state for a low density $\delta $ of
current carriers (doubles or holes in an almost half-filled band).
In particular, Nagaoka proved rigorously the existence of
saturated ferromagnetic state for a single hole at $U\rightarrow
\infty $ and found the instability of the spin-wave spectrum in
the case with increasing $\delta $ and decreasing $U$.

Roth~\cite{Roth:1969} applied a variational principle to this
problem and obtained two critical concentrations. The first one,
$\delta _{\mathrm{c}}$, corresponds to instability of saturated
ferromagnetic state, and the second one, $\delta
_{\mathrm{c}}^{\prime }$, to the transition from non-saturated
ferromagnetic into paramagnetic state. For the simple cubic (sc)
lattice, the values $\delta _{\mathrm{c}}=0.37$ and $\delta
_{\mathrm{c}}^{\prime }=0.64$ were obtained. Next, the region of
stability of the saturated ferromagnetic was investigated within
various approximations in numerous works (see, e.g.,
Refs.~\cite{Jarrett:1968, Sokoloff:1970, Hertz:1973,
Plischke:1974, Takahashi:1982, Vedyaev:1985, Nikolaev:1985,
Zhao:1987, Vedyaev:1988, Ioffe:1988, Fang:1989, Fazekas:1990,
Goryachev:1992, Hanisch:1997, Okabe:1998}).

An interpolational approach to description of magnetic ordering in
narrow bands, which yields saturated ferromagnetism for small
$\delta $ and non-saturated one for large $\delta $, was developed
in Refs.~\cite{Auslender:1988, Auslender:1988a} on the basis of
dynamic magnetic susceptibility treatment. However, the critical
concentrations themselves were not determined.

Using high-temperature expansions in early papers~\cite{Zhao:1987,
Yedidia:1990, Pan:1991, Haaf:1992, Haaf:1995} yielded non-stable
results concerning stability of ferromagnetsm because of poor
accuracy connected with slow convergence~\cite{Plischke:1974}.
However, according to recent results~\cite{Haaf:1992, Haaf:1995},
ferromagnetism also occurs near $\delta =0.3$.

It should be noted that the hole concentration $\delta =1/3$
corresponds to the sign change of the chemical potential in the
Hubbard-I approximation~\cite{Hubbard-I:1963} in the case of
symmetric conduction band, and instability of the paramagnetic
state can occur at this point in some simple
approximations~\cite{Izyumov}.

Experimental data on Fe$_{1-x}$Co$_x$S$_2$~\cite{Jarrett:1968}, a
system with strong correlations, give large values of $\delta
_{\mathrm{c}}$ (saturation ferromagnetism is preserved up to
conduction electron concentrations $n=1-\delta $ of order $0.2$),
but degeneracy effects in the conduction band appear to be
important in this system.

The approaches mentioned do not analyze as a rule in detail the
structure of the one-particle excitation spectrum in the
ferromagnetic phase of the Hubbard model. The simplest
``Hubbard-I'' approximation~\cite{Hubbard-I:1963} for the electron
spectrum corresponds to the zeroth order in the inverse
nearest-neighbor number $1/z$ (``mean-field'' approximation in the
electron transfer). This approximation is quite non-satisfactory
at describing ferromagnetism (in particular, ferromagnetic
solutions are absent, except for peculiar models of bare density
of states).

A consistent calculation of the one-particle Green's functions in
the case of small $\delta $ (almost half-filled band) and low
temperatures was performed in Refs.~\cite{Irkhin:1983,
Irkhin:1985}. The results demonstrated an important role of
non-quasiparticle (incoherent) states in the density-of-states
picture. Expressions for the one-particle Green's functions in a
more wide region of $\delta $ and $T$ were obtained in
Refs.~\cite{Irkhin:1988}.

A physically transparent mechanism of instability of saturated
ferromagnetic state was treated in detail in the works by Edwards
and Hertz~\cite{Hertz:1973}. This mechanism is connected with
occurrence of spin-polaron states above the Fermi level of the
current carriers.

In this paper, the stability of the saturated ferromagnetic state
as the current carrier concentration is raised is studied using
the one-particle Green's functions of first order in~$1/z$ and
corresponding self-consistent approximations. This approach makes
possible to construct a rather simple and physically transparent
picture of the density of states in a saturated Hubbard
ferromagnet. At the same time, the problem of description of
non-saturated state is much more difficult, but our approach turns
out to be successful too.

\section{Calculation of the one-particle Green's functions}

We shall use the Hamiltonian for the Hubbard model in the limit of
infinitely strong Coulomb repulsion in the many-electron
$X$-operator representation~\cite{Hubbard-IV:1965},
\begin{equation}
\mathcal{H}=\sum_{\mathbf{k}\sigma }t_{\mathbf{k}}X_{-\mathbf{k}}^{0\sigma
}X_{\mathbf{k}}^{\sigma 0}
\label{eq:HHM}
\end{equation}
where $t_{\mathbf{k}}$ is the band energy, $X_{\mathbf{k}}^{\alpha
\beta }$ is the Fourier transform of the Hubbard operators
$X_i^{\alpha \beta }=|i\alpha \rangle \langle i\beta |$ ($0$
denotes holes and $\sigma =\pm (\uparrow ,\downarrow )$ denotes
singly occupied states).

It should be noted that in this problem of infinitely strong
Coulomb interaction, a number of difficulties arise in connection
with the non-Fermi excitation statistics. These difficulties occur
both in the diagram technique~\cite{Izyumov} and in the
equations-of-motion method~\cite{Anokhin:1991}. In particular, it
has been found~\cite{Anokhin:1991} that in the expansion with
respect to~$1/z$ the analytic properties of the retarded Green's
functions were violated for the paramagnetic state.

We shall calculate the retarded anticommutator Green's functions
\begin{equation}
G_{\mathbf{k}\sigma }(E)=\langle \!\langle X_{\mathbf{k}}^{\sigma
0}|X_{-\mathbf{k}}^{0\sigma }\rangle \!\rangle _E,\qquad \Im E>0.
\label{eq:EGF}
\end{equation}
We write down the equation of motion
\begin{equation}
(E-t_{\mathbf{k}\sigma })G_{\mathbf{k}\sigma }(E)=(n_0+n_\sigma
)+\Gamma _{\mathbf{k}\sigma }(E),
\label{eq:EM:0}
\end{equation}
where
\[
t_{\mathbf{k}\sigma }=t_{\mathbf{k}}(n_0+n_\sigma ),\qquad
n_\alpha =\langle X_i^{\alpha \alpha }\rangle ,\qquad \alpha =0,\
\sigma ,
\]
\[
\Gamma _{\mathbf{k}\sigma
}(E)=\sum_{\mathbf{q}}t_{\mathbf{k-q}}\langle \!\langle
X_{\mathbf{q}}^{\sigma -\sigma }X_{\mathbf{k-q}}^{-\sigma
0}+\delta (X_{\mathbf{q}}^{00}+X_{\mathbf{q}}^{\sigma \sigma
})X_{\mathbf{k-q}}^{\sigma 0}|X_{-\mathbf{k}}^{0\sigma }\rangle
\!\rangle _E,
\]
with $\delta A=A-\langle A\rangle $. Decoupling the sequence of
equations of motion at the first stage corresponds to the zeroth
order in~$1/z$ and is known as the Hubbard-I approximation. This
may represented in the form
\begin{equation}
G_{\mathbf{k}\sigma }^0(E)=[F_\sigma ^0(E)-t_{\mathbf{k}}]^{-1},
\qquad F_\sigma ^0(E)=\frac E{n_0+n_\sigma }.
\label{eq:GF:0}
\end{equation}
When taking into account fluctuations we obtain
\[
G_{\mathbf{k}\sigma }(E)=G_{\mathbf{k}\sigma }^0(E)\left(
1+\frac{\Gamma _{\mathbf{k}\sigma }(E)}{n_0+n_\sigma }\right) .
\]
Following Ref.~\cite{Irkhin:1988}, we perform decoupling at the
next stage to derive
\[
(E-t_{\mathbf{k-q}-\sigma }-\sigma \omega _{\mathbf{q}})\langle
\!\langle X_{\mathbf{q}}^{\sigma -\sigma
}X_{\mathbf{k-q}}^{-\sigma 0}|X_{-\mathbf{k}}^{0\sigma }\rangle
\!\rangle _E=\chi _{\mathbf{q}}^{\sigma -\sigma
}+n_{\mathbf{k-q}-\sigma }
\]
\begin{equation}
+(t_{\mathbf{k}}\chi _{\mathbf{q}}^{\sigma -\sigma
}-(t_{\mathbf{k-q}}-t_{\mathbf{k}})n_{\mathbf{k-q}-\sigma
})\langle \!\langle X_{\mathbf{k}}^{\sigma
0}|X_{-\mathbf{k}}^{0\sigma }\rangle \!\rangle _E,
\label{eq:EM:1}
\end{equation}
\begin{equation}
(E-t_{\mathbf{k-q}\sigma })\langle \!\langle \delta
(X_{\mathbf{q}}^{00}+X_{\mathbf{q}}^{\sigma \sigma
})X_{\mathbf{k-q}}^{\sigma 0}|X_{-\mathbf{k}}^{0\sigma }\rangle
\!\rangle _E=\chi _{\mathbf{q}}^{-\sigma -\sigma
}(1+t_{\mathbf{k}}\langle \!\langle X_{\mathbf{k}}^{\sigma
0}|X_{-\mathbf{k}}^{0\sigma }\rangle \!\rangle _E),
\label{eq:EM:1f}
\end{equation}
where
\[
\chi _{\mathbf{q}}^{\sigma -\sigma }=\langle S_{\mathbf{q}}^\sigma
S_{-\mathbf{q}}^{-\sigma }\rangle =\langle X_{\mathbf{q}}^{\sigma
-\sigma }X_{\mathbf{-q}}^{-\sigma \sigma }\rangle ,
\]
\[
\chi _{\mathbf{q}}^{-\sigma -\sigma }=\langle \delta
(X_{\mathbf{q}}^{00}+X_{\mathbf{q}}^{\sigma \sigma })\delta
(X_{-\mathbf{q}}^{00}+X_{-\mathbf{q}}^{\sigma \sigma })\rangle
=\langle \delta X_{\mathbf{q}}^{-\sigma -\sigma }\delta
X_{-\mathbf{q}}^{-\sigma -\sigma }\rangle ,
\]
are the correlation function for spin and charge densities
\[
n_{\mathbf{k}\sigma }=\langle X_{\mathbf{-k}}^{0\sigma
}X_{\mathbf{k}}^{\sigma 0}\rangle .
\]
The magnon frequencies in (\ref{eq:EM:1}) are required to cut the
logarithmic ``Kondo'' divergences which are connected with the
Fermi functions.

We write down the Green's function in the locator form
\begin{equation}
G_{\mathbf{k}\sigma }(E)=[F_{\mathbf{k}\sigma
}(E)-t_{\mathbf{k}}]^{-1},\qquad F_{\mathbf{k}\sigma
}(E)=\frac{b_{\mathbf{k}\sigma }(E)}{a_{\mathbf{k}\sigma }(E)}.
\label{eq:GF:1}
\end{equation}
For the Green's function~(\ref{eq:GF:1}) we have
\begin{eqnarray}
a_{\mathbf{k}\sigma }(E) &=&n_0+n_\sigma
+\sum_{\mathbf{q}}t_{\mathbf{k-q}}(\chi _{\mathbf{q}}^{\sigma
-\sigma }+n_{\mathbf{k-q}-\sigma })G_{\mathbf{k-q}-\sigma
}^0(E-\sigma \omega _{\mathbf{q}})
\nonumber \\
&&+\sum_{\mathbf{q}}t_{\mathbf{k-q}}\chi _{\mathbf{q}}^{-\sigma
-\sigma }G_{\mathbf{k-q}\sigma }^0(E),
\label{eq:GF:1:a} \\
b_{\mathbf{k}\sigma }(E)
&=&E+\sum_{\mathbf{q}}t_{\mathbf{k-q}}^2n_{\mathbf{k-q}-\sigma
}G_{\mathbf{k-q}-\sigma }^0(E-\sigma \omega _{\mathbf{q}}).
\label{eq:GF:1:b}
\end{eqnarray}
To simplify our equations, we introduce the magnon spectral
function $K_{\mathbf{q}}(\omega )$ and average this
in~$\mathbf{q}$,
\[
K_{\mathbf{q}}(\omega )=\delta (\omega -\omega
_{\mathbf{q}})\rightarrow \overline{K}(\omega
)=\sum_{\mathbf{q}}K_{\mathbf{q}}(\omega ).
\]
This approximation is sufficient to obtain qualitatively valid
results. Indeed, this approximation (which is in spirit of the
large-$d$ or large-$z$ expansion) reproduces correctly the
low-frequency behavior of spin fluctuations which is important
near the Fermi level. following to Ref.~\cite{Nagaoka:1966} we
have taken below $D=0.7\delta |t|$. It should be noted that the
choice of $D$ influence weakly the critical concentration. We also
neglect $\mathbf{q}$-dependence of transverse and longutidinal
spin correlation functions by taking the values averaged over the
Brillouin zone to obtain
\[
\chi _{\mathbf{q}}^{\sigma -\sigma }=n_\sigma ,\qquad \chi
_{\mathbf{q}}^{-\sigma -\sigma }=n_{-\sigma }(1-n_{-\sigma }).
\]
Then $a(E)$ and $b(E)$ do not depend on~$\mathbf{k}$ and can be
expressed in terms of the bare electron density of states
$N_0(E)$,
\begin{equation}
G_{\mathbf{k}\sigma }(E)=[F_\sigma (E)-t_{\mathbf{k}}]^{-1},\qquad
F_\sigma (E)=\frac{b_\sigma (E)}{a_\sigma (E)},
\label{eq:GF:1:K}
\end{equation}
\begin{eqnarray}
a_\sigma (E) &=&n_0+n_\sigma +\int \overline{K}(\omega
)\sum_{\mathbf{q}}t_{\mathbf{q}}(n_\sigma +n_{\mathbf{q}-\sigma
})G_{\mathbf{q}-\sigma }^0(E-\sigma \omega )d\omega
\nonumber \\
&&+\sum_{\mathbf{q}}t_{\mathbf{q}}n_{-\sigma }(1-n_{-\sigma
})G_{\mathbf{q}\sigma }^0(E),
\label{eq:GF:1:K:a} \\
b_\sigma (E) &=&E+\int \overline{K}(\omega
)\sum_{\mathbf{q}}t_{\mathbf{q}}^2n_{\mathbf{q}-\sigma
}G_{\mathbf{q}-\sigma }^0(E-\sigma \omega )d\omega ,
\label{eq:GF:1:K:b}
\end{eqnarray}
\[
G_{\mathbf{q}-\sigma }^0(E-\sigma \omega )\rightarrow G_{-\sigma
}^0(E-\sigma \omega ,t)=[F_\sigma (E-\sigma \omega
)-t]^{-1},\qquad F_\sigma (E)=\frac E{n_0+n_{-\sigma }},
\]
\begin{eqnarray*}
a_\sigma (E) &=&n_0+n_\sigma +\int \overline{K}(\omega )\int
N_0(t)t(n_\sigma +f_{-\sigma }(t))G_{-\sigma }^0(E-\sigma \omega
,t)d\omega dt
\\
&&+\int N_0(t)tn_{-\sigma }(1-n_{-\sigma })G_\sigma ^0(E,t)dt,
\\
b_\sigma (E) &=&E+\int \overline{K}(\omega )\int
N_0(t)t^2f_{-\sigma }(t)G_{-\sigma }^0(E-\sigma \omega ,t)d\omega
dt,
\end{eqnarray*}
where $f_\sigma (t_{\mathbf{q}})=n_{\mathbf{q}\sigma }$,
$N_0(t)=\sum_{\mathbf{k}}\delta (t-t_{\mathbf{k}})$ is the bare
density of states. The exact quasiparticle density of states is
given by
\begin{equation}
N_\sigma (E)=-\frac 1\pi \Im \sum_{\mathbf{k}}G_{\mathbf{k}\sigma
}(E).
\label{eq:DOS}
\end{equation}
\[
N_\sigma (E)=-\frac 1\pi \Im \int\limits_{-\infty }^{+\infty
}G_\sigma (E,t)N_0(t)dt.
\]

Now we write down the self-consistent approximation with replacing
in (\ref{eq:GF:1:K:a}), (\ref{eq:GF:1:K:b}) the bare locator
$F_\sigma ^0(E)$ by the exact locator $F_\sigma (E)$, i.e.
\begin{equation}
G_{\mathbf{k}\sigma }^0(E)\rightarrow G_{\mathbf{k}\sigma }(E).
\label{eq:s-c}
\end{equation}
In such an approach, large electron damping is present which
smears the ``Kondo'' peak, so that including magnon frequencies
does not change qualitatively the picture, but may be important
for quantitatative results. We obtain
\begin{eqnarray}
G_{\mathbf{k}\sigma }(E) &=&[F_\sigma
(E)-t_{\mathbf{k}}]^{-1},\qquad F_\sigma (E)=\frac{B_\sigma
(E)}{A_\sigma (E)},
\label{eq:GF:1:s-c} \\
A_\sigma (E) &=&n_0+n_\sigma +\frac 1{n_0+n_{-\sigma }}\int
\overline{K}(\omega )\sum_{\mathbf{q}}t_{\mathbf{q}}(n_\sigma
+n_{\mathbf{q}-\sigma })G_{\mathbf{q}-\sigma }(E-\sigma \omega
)d\omega
\nonumber \\
&&\ +\frac 1{n_0+n_\sigma
}\sum_{\mathbf{q}}t_{\mathbf{q}}n_{-\sigma }(1-n_{-\sigma
})G_{\mathbf{q}\sigma }(E),
\label{eq:GF:1:s-c:a} \\
B_\sigma (E) &=&E+\frac 1{n_0+n_{-\sigma }}\int
\overline{K}(\omega
)\sum_{\mathbf{q}}t_{\mathbf{q}}^2n_{\mathbf{q}-\sigma
}G_{\mathbf{q}-\sigma }(E-\sigma \omega )d\omega .
\label{eq:GF:1:s-c:b}
\end{eqnarray}
\begin{eqnarray*}
A_\sigma (E) &=&n_0+n_\sigma +\frac 1{n_0+n_{-\sigma }}\int
\overline{K}(\omega )\int N_0(t)t(n_\sigma +f_{-\sigma
}(t))G_{-\sigma }(E-\sigma \omega ,t)dtd\omega
\\
&&\ +\frac 1{n_0+n_\sigma }\int N_0(t)tn_{-\sigma }(1-n_{-\sigma
})G_\sigma (E,t)dt,
\\
B_\sigma (E) &=&E+\frac 1{n_0+n_{-\sigma }}\int
N_0(t)t^2f_{-\sigma }(t)G_{-\sigma }(E-\sigma \omega ,t)dtd\omega
.
\end{eqnarray*}
It should be noted that another self-consistent approximation used
in Refs.~\cite{Anokhin:1991} leads to not quite satisfactory
results because of violation of normalization condition for the
density of states. As demonstrate our calculations, such a
difficulty exists also for the approximation (\ref{eq:GF:1:K}),
but the violation is numerically small for $\delta <\delta
_{\mathrm{c}}$ (about 2\%); note that introducing longitudinal
fluctuations improves considerably the results in comparison with
Ref.~\cite{Zarubin:1999}. At the same time, the approximation
(\ref{eq:GF:1:s-c}) with the locator structure of the Green's
function does not violate the analytical properties.

The chemical potential $\mu $ is determined by the number of
holes. By using the spectral representation for the Green's
function (\ref{eq:EGF}) this is calculated as
\begin{equation}
\delta \equiv n_0=\langle X^{00}\rangle =\langle X_i^{0\sigma
}X_i^{\sigma 0}\rangle =\sum\limits_{\mathbf{k}}\langle
X_{\mathbf{k}}^{0\sigma }X_{-\mathbf{k}}^{\sigma 0}\rangle =-\frac
1\pi \Im \sum_{\mathbf{k}}\int\limits_{-\infty }^{+\infty
}G_{\mathbf{k}\sigma }(E)f(E)dE.
\label{eq:n0}
\end{equation}
It is important the Hubbard-I approximation is hardly satisfactory
in the narrow-band ferromagnetism problem since it is difficult to
formulate a reasonable criterion of magnetic ordering by direct
using the expressions for one-electron Green's functions like
(\ref{eq:GF:0}). Unlike the decoupling scheme by Hubbard
\cite{Hubbard-I:1963}, the many-electron $X$-operator approach
clarifies the causes of this failure. In particular, one can see
that the approximation (\ref{eq:GF:0}) violates the kinematical
requirements since it is impossible to satisfy at $\langle
S^z\rangle \neq 0$ the relation (\ref{eq:n0}) for both spin
projections $\sigma $. Indeed, the quasiparticle pole for~$\sigma
=\downarrow $, corresponding to a narrowed band and lying above
the Fermi level of the holes, does not provide an adequate
description of the energy spectrum and leads to the appearance of
finite~$n_{\downarrow }$, i.e., the saturation ferromagnetism
cannot be properly treated. However, the situation changes
provided we use the expressions containing first-order
$1/z$-corrections. Unlike the Hubbard-I approximation, the value
of $\mu $ turns out to be weakly dependent of $\sigma $ for the
approximation (\ref{eq:GF:1:K}) and independent of $\sigma $ for
the the approximation (\ref{eq:GF:1:s-c}).

The magnetization is determined from the equation
\begin{equation}
\langle S^z\rangle =\frac 12\sum\limits_\sigma \sigma n_\sigma ,
\label{eq:magn}
\end{equation}
with
\[
n_\sigma =\sum_{\mathbf{k}}\langle X_{-\mathbf{k}}^{\sigma
0}X_{\mathbf{k}}^{0\sigma }\rangle =\int\limits_{-\infty
}^{+\infty }N_\sigma (E)(1-f(E))dE.
\]
In the leading approximation in~$1/z$ for the one-particle
occupation numbers, it is neccessary to use the Hubbard-I
approximation, i.e.,
\[
n_{\mathbf{k}\sigma }=(n_0+n_\sigma )f(t_{\mathbf{k}\sigma }),
\]
but the chemical potential should be already chosen from the
Green's function~(\ref{eq:GF:1}). As opposed to
Eq.~(\ref{eq:GF:0}), the Green's functions~(\ref{eq:GF:1}) contain
terms with resolvents and have branch cuts which describe
non-quasiparticle (incoherent) contributions to the density of
states. It is the latter which ensure qualitative agreement with
the sum rule~(\ref{eq:n0}) for~$\sigma =\downarrow $. At the same
time, there are no poles of the Green's function for this
projection of the spin for small $\delta $ above the Fermi level,
i.e., the saturated ferromagnetic state is preserved.

The full density of states can be also represented in terms of the
exact resolvent $R_\sigma (E)=\sum_{\mathbf{k}}G_{\mathbf{k}\sigma
}(E)$ as
\[
N_\sigma (E)=-\frac 1\pi \Im R_\sigma (E).
\]
This quantity satisfies the equation
\[
R_\sigma (E)=R_0\{F_\sigma (E)\},\qquad
R_0(E)=\sum_{\mathbf{k}}\frac 1{E-t_{\mathbf{k}}}=\int N_0(t)\frac
1{E-t}dt.
\]

In the case of a saturated ferromagnetic state, the Green's
function~(\ref{eq:GF:1}) takes the form
\begin{eqnarray}
G_{\mathbf{k}\uparrow }(E) &=&\left( E\left/ \left[
1+\sum\limits_{\mathbf{q}}\frac{t_{\mathbf{q}}(1-n_0)}{E-t_{\mathbf{q}}n_0}\right]
-t_{\mathbf{k}}\right. \right) ^{-1},
\label{eq:GF:EH:up} \\
G_{\mathbf{k}\downarrow }(E)
&=&E\sum_{\mathbf{q}}\frac{n_{\mathbf{k-q}}}{E-t_{\mathbf{k-q}}+\omega
_{\mathbf{q}}}
\nonumber \\
&&\times \left( E\left[
1-n_0+\sum_{\mathbf{q}}\frac{(E-t_{\mathbf{k}})n_{\mathbf{k-q}}}{E-t_{\mathbf{k-q}}+\omega
_{\mathbf{q}}}\right]
-\sum_{\mathbf{q}}t_{\mathbf{k-q}}n_{\mathbf{k-q}}\right) ^{-1},
\label{eq:GF:EH:down}
\end{eqnarray}
where $n_{\mathbf{k}}=f(t_{\mathbf{k}})$. Note that, as opposed to
the one-electron approach~\cite{Hertz:1973}, the Green's
function~$G_{\mathbf{k}\uparrow }(E)$ does not reduce to the free
electron Green's function even in the saturated ferromagnetic
state, since fluctuations in the hole occupation number contribute
to it. Neglecting the resolvent in Eq.~(\ref{eq:GF:EH:up}) and the
last term in the denominator of Eq.~(\ref{eq:GF:EH:down}), we
obtain a somewhat different form of the Green's function in terms
of the electron self-energy
\begin{equation}
G_{\mathbf{k}\uparrow }(E)=\frac 1{E-t_{\mathbf{k}}},\qquad
G_{\mathbf{k}\downarrow }(E)=\frac 1{E-t_{\mathbf{k}}-\Sigma
_{\mathbf{k}\downarrow }(E)},
\label{eq:GF:EH}
\end{equation}
\[
\Sigma _{\mathbf{k}\downarrow }(E)=-(1-n_0)\left(
\sum_{\mathbf{q}}\frac{n_{\mathbf{k-q}}}{E-t_{\mathbf{k-q}}+\omega
_{\mathbf{q}}}\right) ^{-1}.
\]
This result corresponds to the Edwards-Hertz approximation in the
limit $U\rightarrow \infty $. Of course, these expressions work
only in the saturated ferromagnetic state. The results
(\ref{eq:GF:EH}) can be also obtained by using the expansion in
the electron and magnon occupation numbers~\cite{Irkhin:1990}.

\section{Results of calculations and discussion}

\begin{figure}[htbr]
\begin{center}
\includegraphics[width=0.5\textwidth, angle=-90]{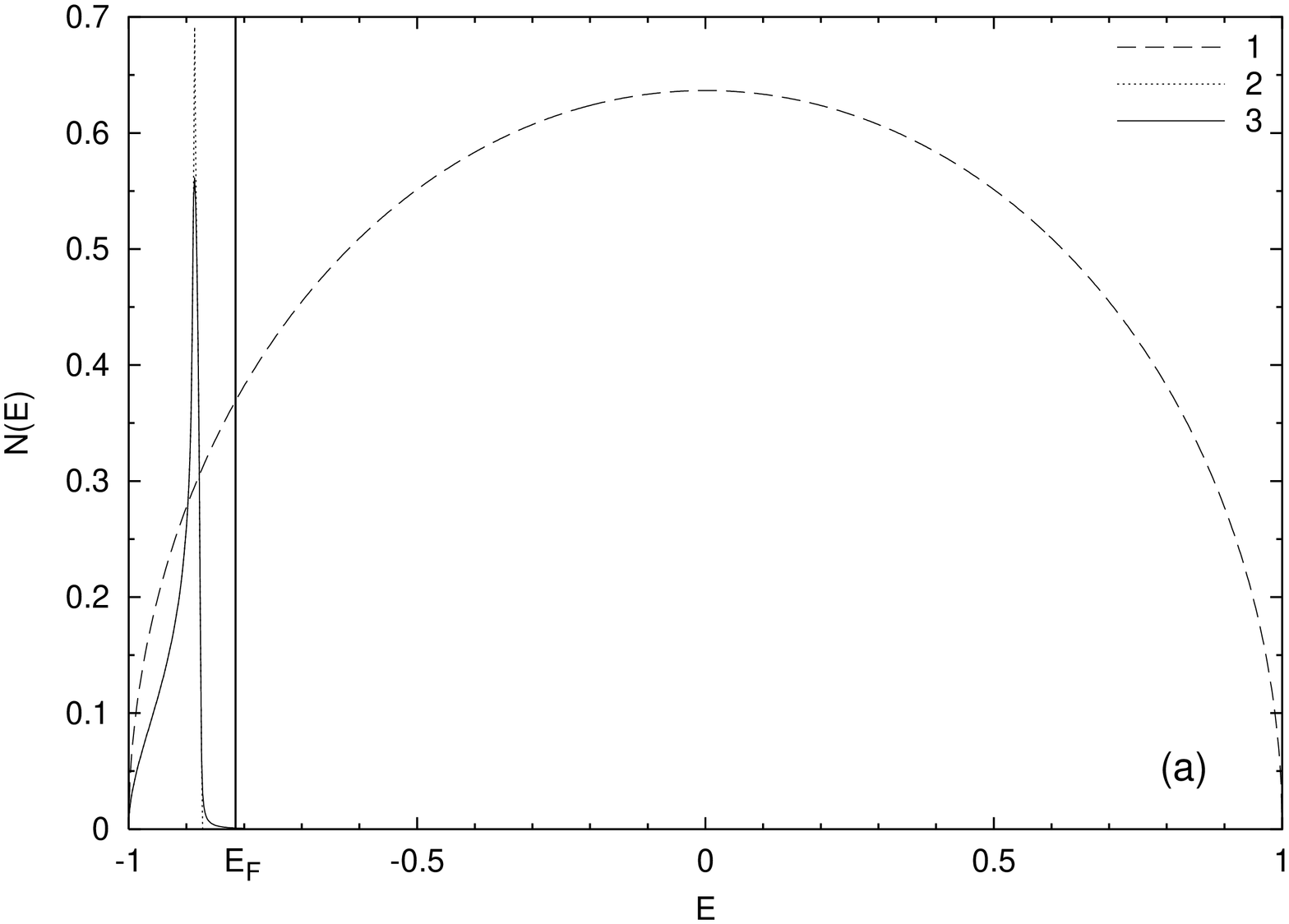}
\includegraphics[width=0.5\textwidth, angle=-90]{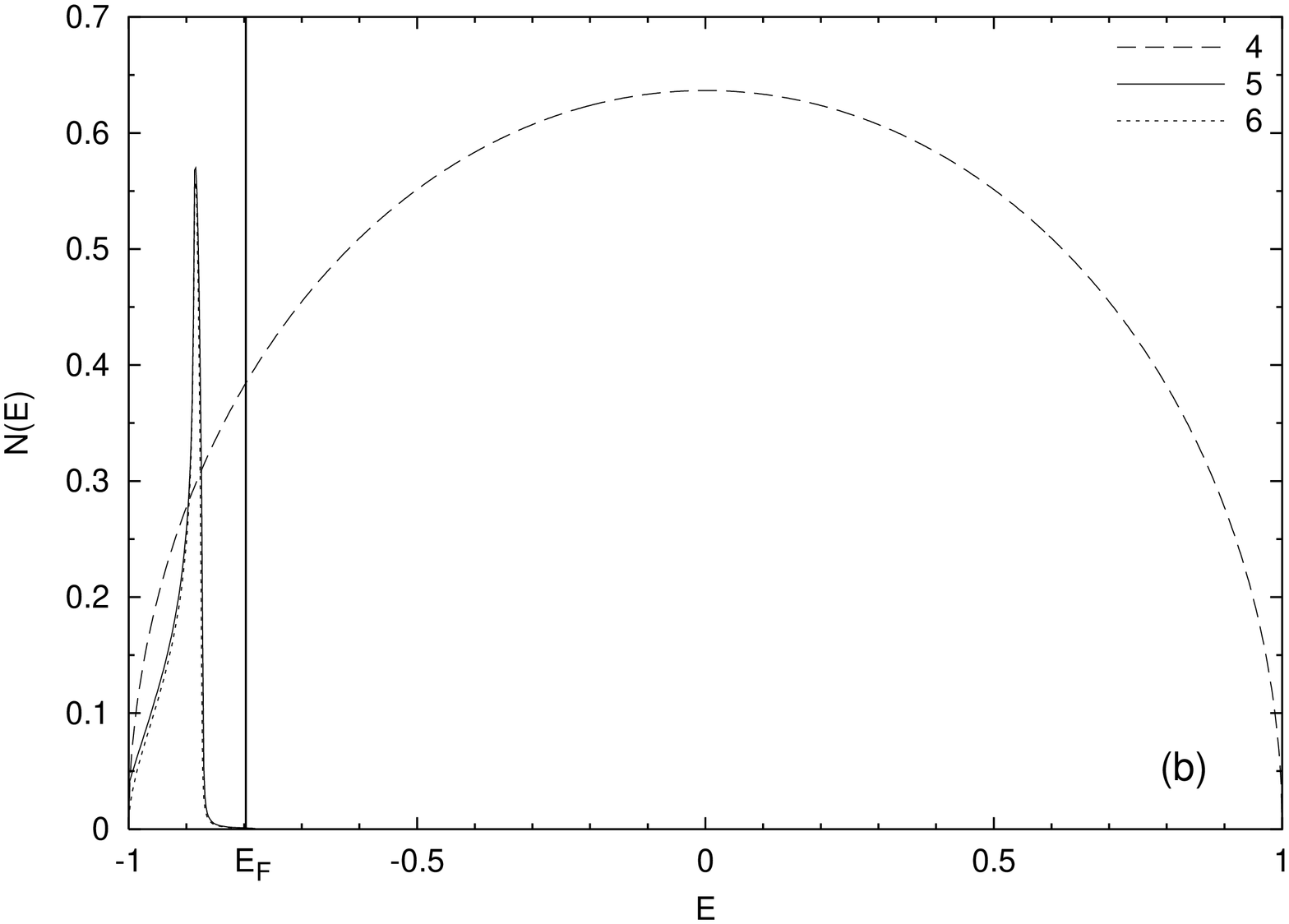}
\end{center}
\caption{Density of states for the semielliptic DOS at
concentration of carriers current~$\delta =0.02$. (a) line~1
($\sigma =\uparrow $) and~line~2 ($\sigma =\downarrow $)
correspond to the non-self-consistent
approximation~(\ref{eq:GF:1}); line~1 ($\sigma =\uparrow $)
and~line~3 ($\sigma =\downarrow $) to the non-self-consistent
approximation with account of spin dynamics~(\ref{eq:GF:1:K}); (b)
line~4 ($\sigma =\uparrow $) and~line~5 ($\sigma =\downarrow $)
correspond to the self-consistent
approximation~(\ref{eq:GF:1:s-c}), and~line 4 ($\sigma =\uparrow
$) and line~6 ($\sigma =\downarrow $) to the Edwards-Hertz
approximation~(\ref{eq:GF:EH}).}
\label{fig:1}
\end{figure}

The $1/z$ corrections lead to a non-trivial structure of the total
quasiparticle density of states. In the non-self-consistent
approach the integral with the Fermi functions yields, similar to
the Kondo problem, the logarithmic singularity
\[
\sum_{\mathbf{q}}\frac{f(t_{\mathbf{k+q}})}{E-t_{\mathbf{k+q}}}\simeq
-\ln |E-E_{\mathrm{F}}|N(E_{\mathrm{F}}).
\]
For very low $\delta $ a significant logarithmic singularity
exists only in the imaginary part of the Green's function, which
corresponds to a finite jump in the density of
states~\cite{Irkhin:1983, Irkhin:1985}. However, when $\delta $
increases, it is necessary to take into account the resolvents in
both the numerator and denominator of the Green's function, so
that the real and imaginary parts are ``mixed'' and a logarithmic
singularity appears in the density of states. When the magnon
frequencies are included in the denominators of
Eqs.~(\ref{eq:GF:1:K:a}) and~(\ref{eq:GF:1:K:b}), the singularity
is spread out over the interval~$\omega _{\mathrm{max}}$ and the
peak is smoothed out. In the self-consistent
approximations~(\ref{eq:GF:1:s-c:a}) and~(\ref{eq:GF:1:s-c:b}) the
form of~$N_{\downarrow }(E)$ approaches the bare density of states
and the peak is completely smeared, even neglecting spin
dynamics~(Fig.~\ref{fig:1}), so that the latter plays no crucial
role, although shifts somewhat the peak below the Fermi level.

\begin{figure}[htbr]
\begin{center}
\includegraphics[width=0.5\textwidth, angle=-90]{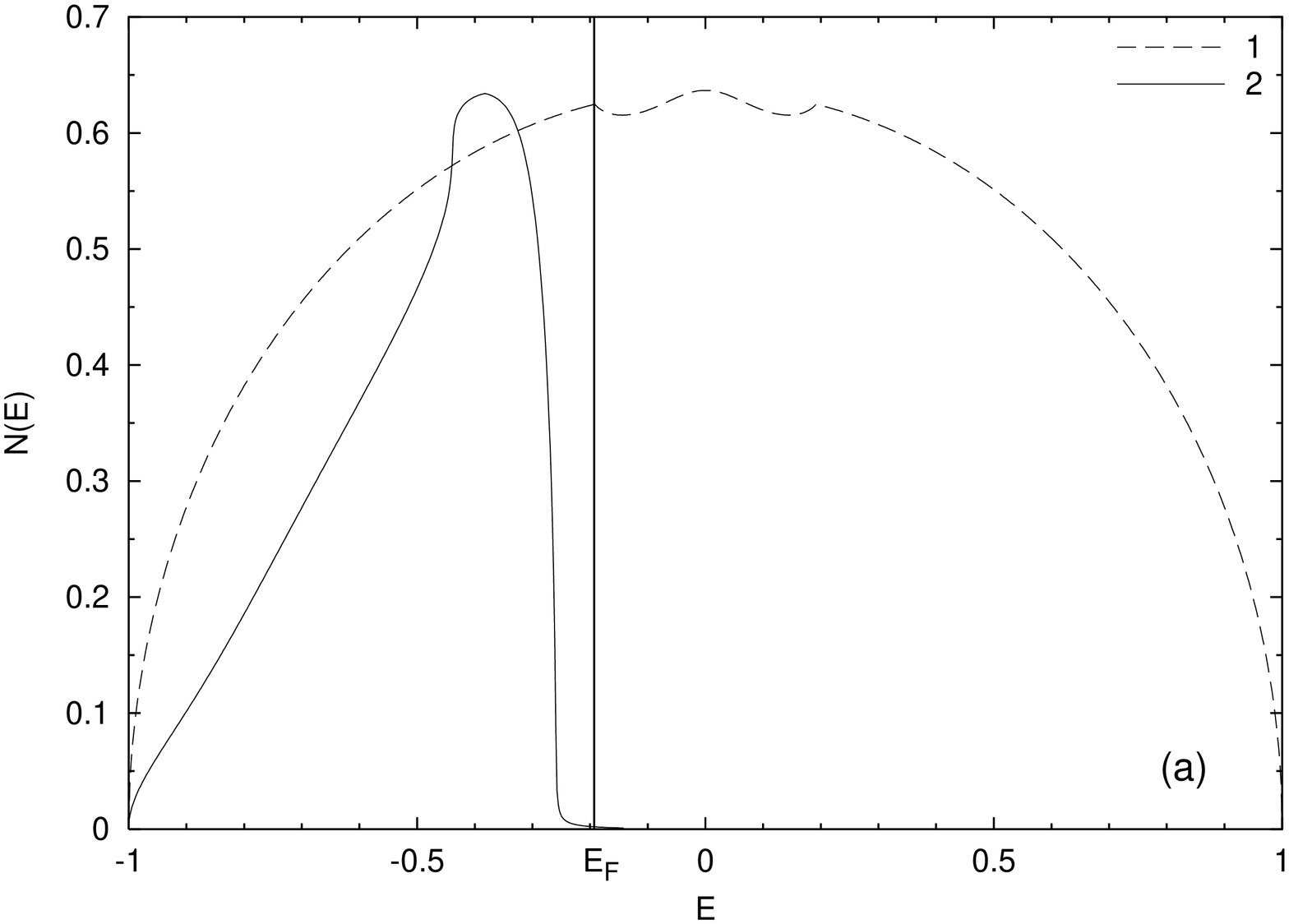}
\includegraphics[width=0.5\textwidth, angle=-90]{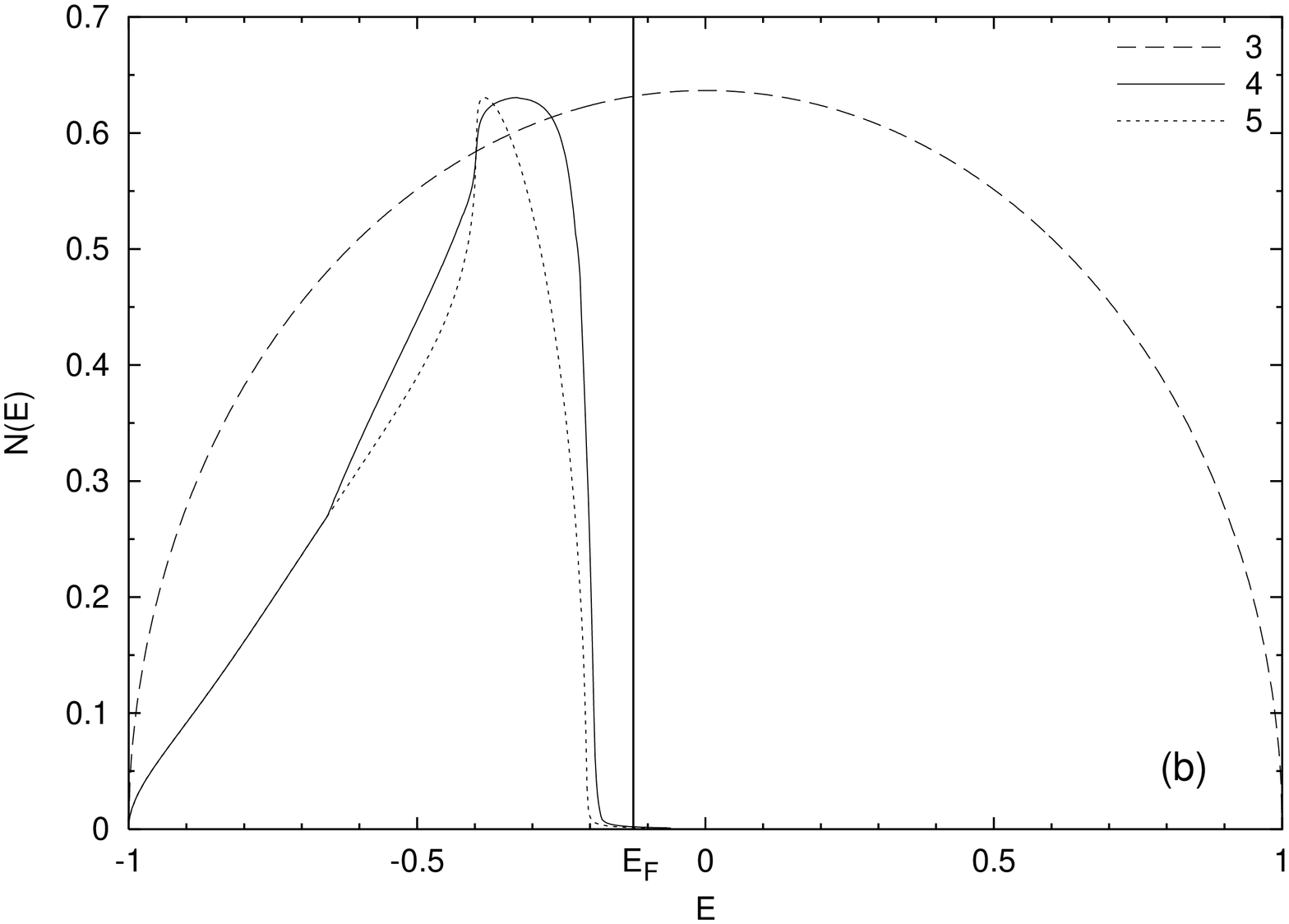}
\end{center}
\caption{Density of states for the semielliptic DOS at
concentration of carriers current~$\delta =0.20$. (a) line~1
($\sigma =\uparrow $) and~line~2 ($\sigma =\downarrow $)
correspond to the non-self-consistent approximation with spin
dynamics~(\ref{eq:GF:1:K}); (b) line~3 ($\sigma =\uparrow $)
and~line~4 ($\sigma =\downarrow $) correspond to the
self-consistent approximation~(\ref{eq:GF:1:s-c}), and~line 3
($\sigma =\uparrow $) and line~5 ($\sigma =\downarrow $) to the
Edwards-Hertz approximation~(\ref{eq:GF:EH}).}
\label{fig:2}
\end{figure}

Near the critical concentration the peak in
approximation~(\ref{eq:GF:1}) (but not in the Edwards-Hertz
approximation) is again smeared~(Fig.~\ref{fig:2}), but this
spreading out is no longer noticeable for~$\delta =0.15$.

\begin{figure}[htbr]
\begin{center}
\includegraphics[width=0.5\textwidth, angle=-90]{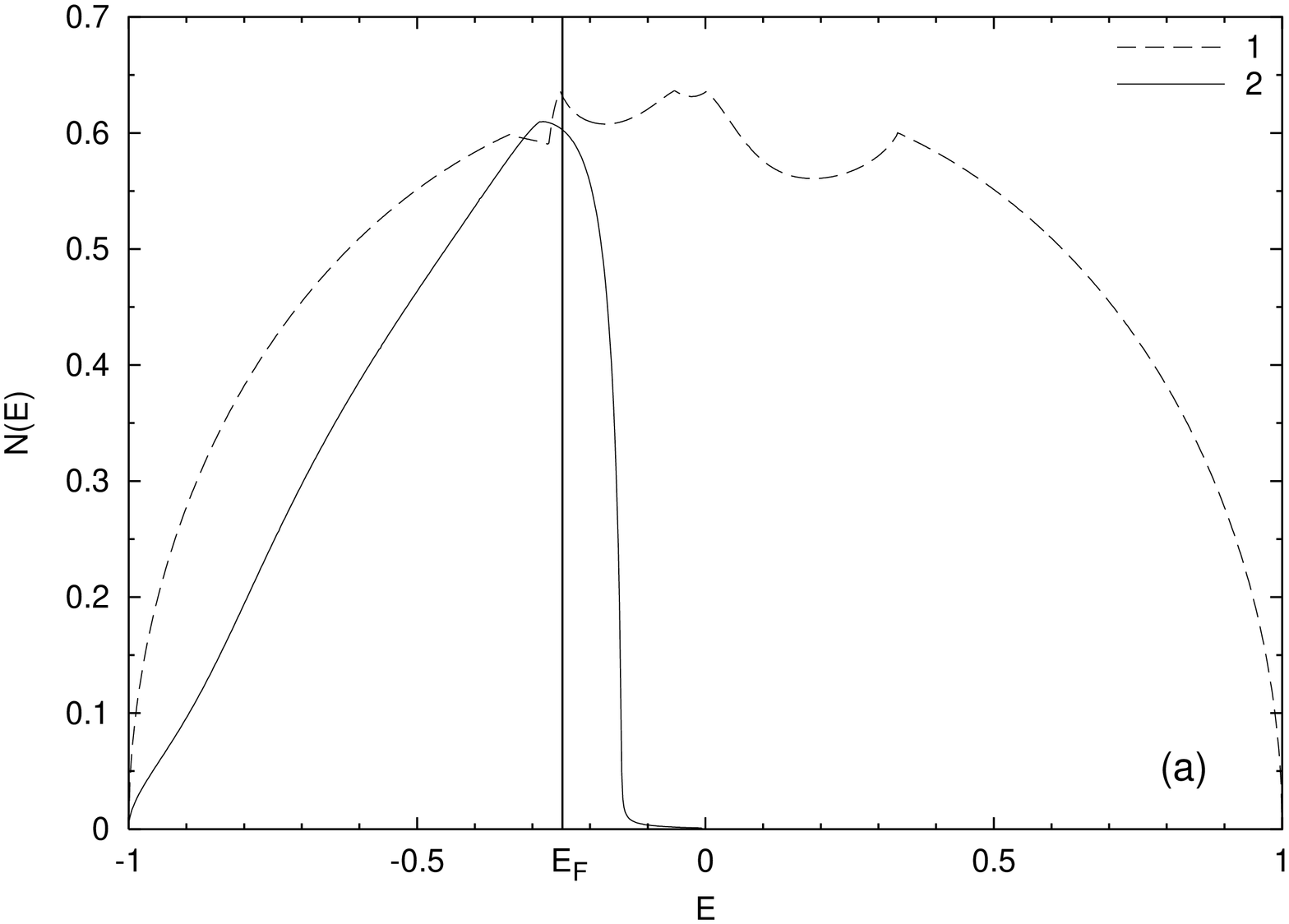}
\includegraphics[width=0.5\textwidth, angle=-90]{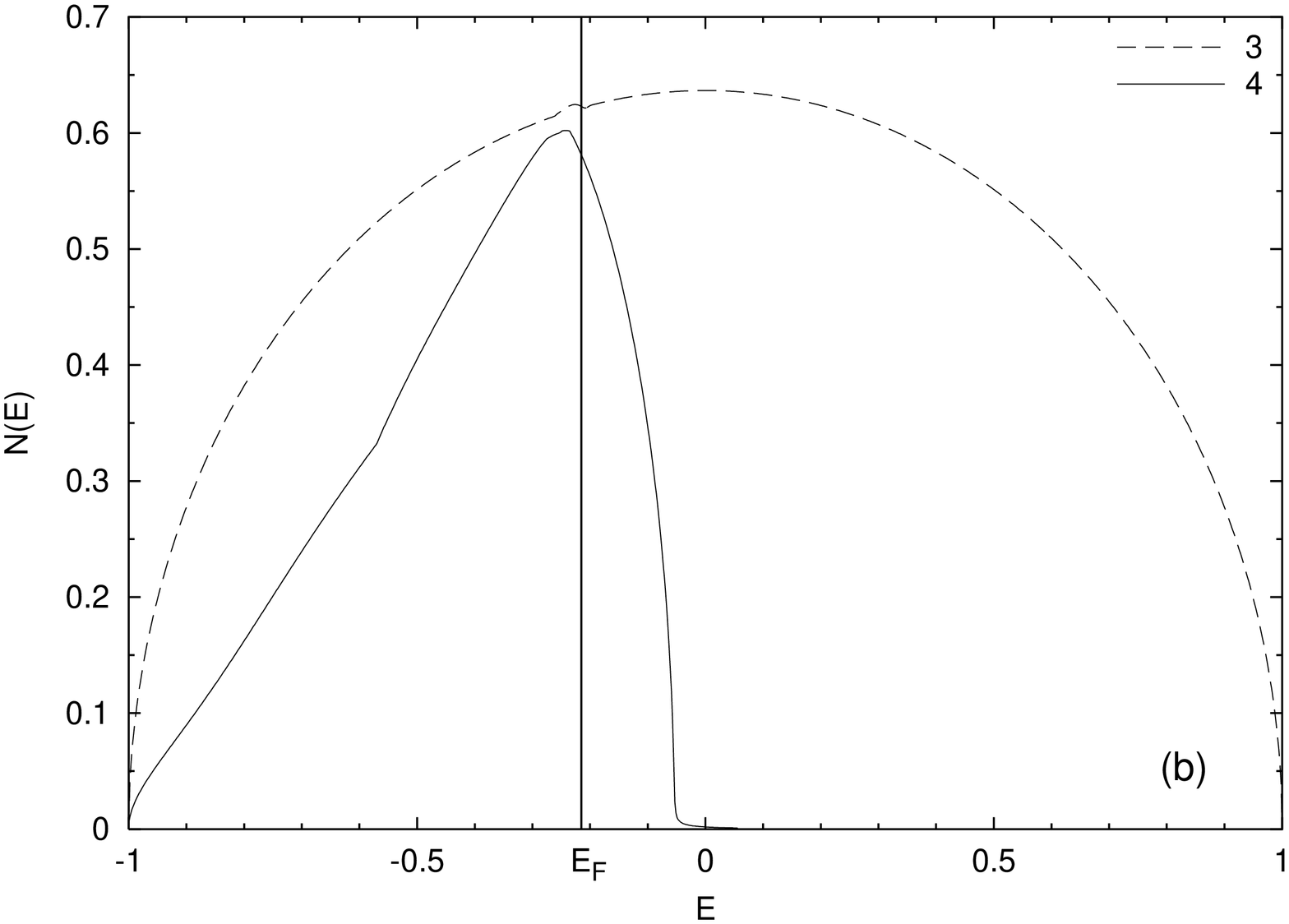}
\end{center}
\caption{Density of states for the semielliptic DOS at
concentration of carriers current~$\delta =0.35$. (a) line~1
($\sigma =\uparrow $) and~line~2 ($\sigma =\downarrow $)
correspond to the non-self-consistent approximation with spin
dynamics~(\ref{eq:GF:1:K}); (b) line~3 ($\sigma =\uparrow $)
and~line~4 ($\sigma =\downarrow $) correspond to the
self-consistent approximation~(\ref{eq:GF:1:s-c}).}
\label{fig:3}
\end{figure}

In the non-saturated state a spin-polaron pole occurs, so that
quasiparticle states with $\sigma =\downarrow $ occur above the
Fermi level with
\[
n_{\downarrow }=\int dEf(E)N_{\downarrow }(E)
\]
The corresponding DOS picture is shown in Fig.~\ref{fig:3}.

The critical concentrations~$\delta _{\mathrm{c}}$ for the loss of
stability of saturated ferromagnetism, as calculated in different
approximations considered, are listed in Table~\ref{tab:1} for a
number of bare densities of states. In the case of fcc lattices
(where the bare density of states is asymmetric and has a
logarithmic divergence on one edge) we have chosen the sign of the
transfer integral for which the saturated ferromagnetism is stable
at low $\delta $~\cite{Nagaoka:1966}. Note that it is necessary to
use an equation for the chemical potential~(\ref{eq:n0}) that is
derived from the complete Green's functions~(\ref{eq:GF:1:s-c}).
(Using the Hubbard-I approximation here leads to a drop in~$\delta
_{\mathrm{c}}$ of the order of $0.1$.)

It is clear from the Table~\ref{tab:1} that the results are fairly
stable and do not depend too strong on the form of the
approximation. In particular, self-consistency changes them
little, leading to a slight reduction in~$\delta _{\mathrm{c}}$.
The dependence on spin dynamics (magnon spectrum), even in the
non-selfconsistent approximation, is weaker (the critical
concentration only varies in the third decimal place). At the same
time spin dymamics is important for the description of the states
near the Fermi level. It is interesting to note that results of
the Edwards-Hertz approximation~(\ref{eq:GF:EH}) are closer to
those of the self-consistent approximation~(\ref{eq:GF:1:s-c})
than of the non-self-consistent approximation~(\ref{eq:GF:1}).
Unfortunately, in Ref.~\cite{Hertz:1973} only a crude estimate
of~$\delta _{\mathrm{c}}$ was made by using the quadratic
dispersion relation for the hole spectrum which yielded $\delta
_{\mathrm{c}}=0.16$. This approximation is not sufficient for
quantitative calculations, as one can see from Table~\ref{tab:1}.

Unlike most other analytical approaches, our results for the
one-particle Green's describe formation of non-saturated
ferromagnetism too. It is important that the account of
longitudinal spin fluctuations (which were neglected in
Ref.~\cite{Zarubin:1999}) turns out to be important for obtaining
the non-saturated solution and calculating the second critical
concentration $\delta _{\mathrm{c}}^{\prime }$

\begin{figure}[htbr]
\begin{center}
\includegraphics[width=0.7\textwidth, angle=-90]{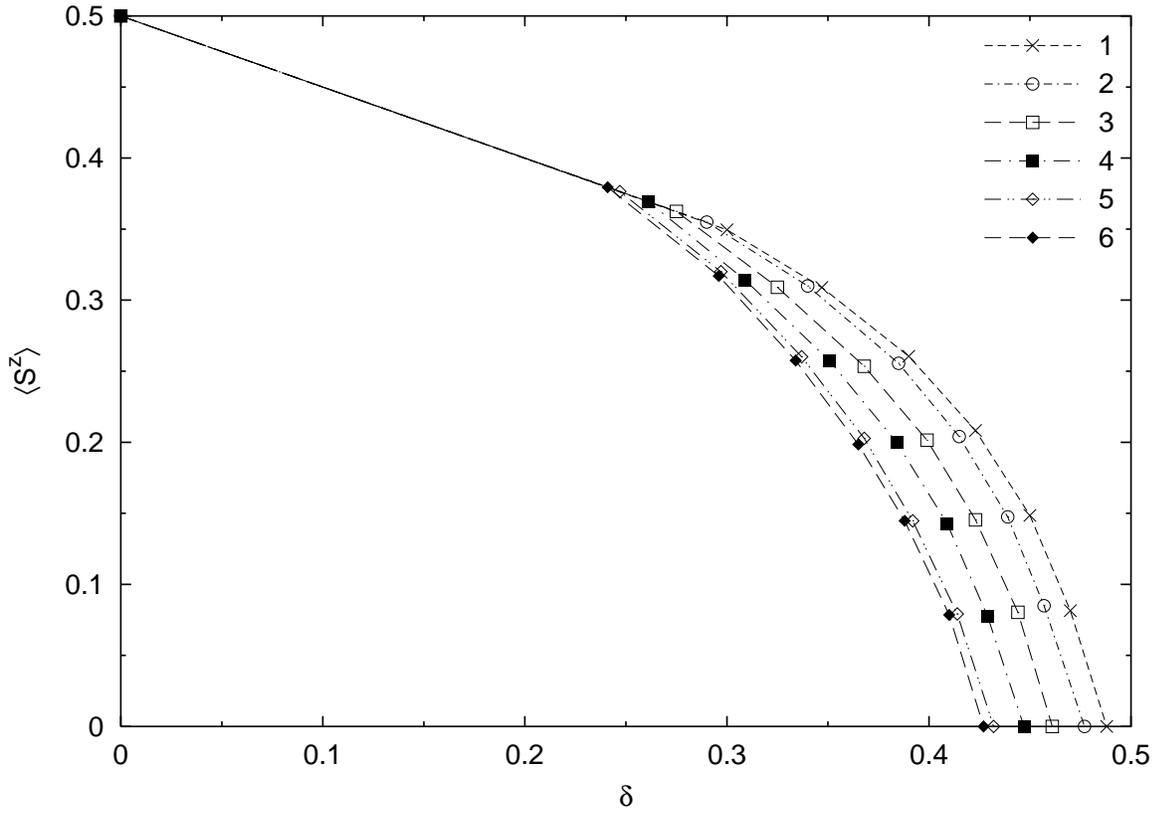}
\end{center}
\caption{The dependence of the magnetization~$\langle S^z\rangle $
on concentration of current carriers~$\delta $ for a number of
bare DOS's. Line~1 corresponds to rectangular DOS, line~2~to
semielliptic DOS, lines~3, 4, 5, 6~to square, simple cubic (sc),
bcc, and fcc latices, respectively.}
\label{fig:4}
\end{figure}

The dependence of the saturation magnetization on the
concentration of current carriers for various bare DOS's is shown
in Fig.~\ref{fig:4}. One can see that this dependence deviates
from the linear one, $\langle S^z\rangle = (1-n)/2$, for $\delta >
\delta_{\mathrm{c}}$.

Let us perform a comparison of our results with other
calculations. Generally, most calculations for a number of
lattices yield the value of $\delta _{\mathrm{c}}$ which is close
to $0.3$ (although the small value $\delta _{\mathrm{c}}=0.045$
was obtained in Ref.~\cite{Nikolaev:1985} for sc lattice by a
diagram approach, a close approach of Ref.~\cite{Ioffe:1988}
yields much larger values, e.g., $\delta _{\mathrm{c}}=0.25$ for
the quadratic lattice). At the same time, for the critical
concentrations~$\delta _{\mathrm{c}}^{\prime }$ the interval of
values is broader and varies from~$0.38$ to~$0.64$. Our
calculations yield the $\delta _{\mathrm{c}}^{\prime }$ values
which are considerably smaller than the results of the spin-wave
approximation \cite{Roth:1969}.

Improved Gutzwiller method~\cite{Fazekas:1990} yields for the sc
lattice $\delta _{\mathrm{c}}=0.33$, and using the $t/U$
expansion~\cite{Zhao:1987} yields $\delta _{\mathrm{c}}=0.27$. For
the quadratic lattice the result of the variational
approach~\cite{Wurth:1996} is $\delta _{\mathrm{c}}=0.251$, and
the result of Ref.~\cite{Kotliar:1986} is $\delta
_{\mathrm{c}}^{\prime }=0.38$. The density matrix renormalization
group approach~\cite{Liang:1995} lead to the value $\delta
_{\mathrm{c}}=0.22$ and the rough estimate $\delta
_{\mathrm{c}}^{\prime }\simeq 0.40$. Quantum Monte-Carlo (QMC)
method for density matrix in Ref.~\cite{Liang:1995} gives $\delta
_{\mathrm{c}}=0.22$ and rough evaluation~$\delta
_{\mathrm{c}}^{\prime }\approx 0.40$. QMC in 2d
case~\cite{Becca:2001} gives~$\delta _{\mathrm{c}}^{\prime
}\approx 0.40$. Self-consistent spin density
approximation~(SDA)~\cite{Herrmann:1997} lead to the results for
simple cubic and bcc lattices~$\delta _{\mathrm{c}}<0.32;$ the
values of $\delta _{\mathrm{c}}^{\prime }$ for the cubic lattices
are given in Table~\ref{tab:1}.

The values we have obtained can be compared with those in the
limit of an infinite-dimensional space (it should be expected that
our method of expanding in powers of~$1/z$ is rather close to this
approximation), for which $\delta _{\mathrm{c}}=0.42$
(Ref.~\cite{Fazekas:1990}) and $\delta _{\mathrm{c}}=0.33$
(Ref.~\cite{Obermeier:1997}) have been obtained. At the same time,
our approach is possible to reproduce the dependence of~$\delta
_{\mathrm{c}}$ on the dimensionality of space and on the form of
the bare density of states.

Recently, $\delta _{\mathrm{c}}$ has been obtained for a large
number of lattices~\cite{Shastry:1990, Hanisch:1997}. These
results are also given in the Table~\ref{tab:1} for comparison. It
can be seen that in a number of cases our results agree better
with a number of other calculations, especially for a square
lattice. We note in this connection that a variational method has
been used~\cite{Linden:1991} to obtain a rigorous estimate of
$\delta _{\mathrm{c}}<0.29$ for a square lattice. Therefore, our
results can be regarded as fairly reliable, even quantitatively.


To conclude, we have obtained the density-of-states pictures in a
Hubbard ferromagnet with account of the ``Kondo'' scattering and
spin-polaron contributions. Our approach yield a rather simple
interpolational description of saturated and non-saturated
ferromagnetism. One can expect that the results obtained will be
useful for qualitative understanding of the ferromagnetism
formation in narrow bands.

The research described was supported in part by Grant
No.~747.2003.2 (Support of Scientific Schools) from the Russian
Basic Research Foundation.

\begin{table}[htbr]
\caption{Values of critical concentrations $\delta _{\mathrm{c}}$
and $\delta _{\mathrm{c}}^{\prime }$ for rectangular (ra) and
semielliptic (se) bare density of state, square, simple cubic
(sc), bcc, fcc lattices. I, VII~is non-self-consistent
approximation~(\ref{eq:GF:1}), II is Edwards-Hertz
approximation~(\ref{eq:GF:EH}), III, VIII~is self-consistent
approximations (with fluctuations)~(\ref{eq:GF:1:s-c}), IV~is
results of work~\cite{Shastry:1990}, V, IX~is results of
work~\cite{Herrmann:1997} VI~is the result of
Ref.~\cite{Hanisch:1997} (variant of calculation RES0, for fcc
lattices instability not discovered).}
\label{tab:1}
\begin{center}
\begin{tabular}{llllllllll}
\hline
\multicolumn{1}{l|}{DOS} & \multicolumn{6}{c|}{$\delta _{\mathrm{c}}$} &
\multicolumn{3}{c}{$\delta _{\mathrm{c}}^{\prime }$} \\ \cline{2-10}
\multicolumn{1}{l|}{} & I & II & III & IV & V & \multicolumn{1}{l|}{VI} & VII
& VIII & IX \\ \hline
\multicolumn{1}{l|}{ra} & $0.276$ & $0.284$ & $0.301$ &  &  &
\multicolumn{1}{l|}{} & $0.468$ & $0.488$ &  \\
\multicolumn{1}{l|}{se} & $0.258$ & $0.266$ & $0.290$ &  &  &
\multicolumn{1}{l|}{} & $0.458$ & $0.477$ &  \\
\multicolumn{1}{l|}{square} & $0.244$ & $0.252$ & $0.275$ & $0.49$ &  &
\multicolumn{1}{l|}{$0.4045$} & $0.449$ & $0.461$ &  \\
\multicolumn{1}{l|}{sc} & $0.233$ & $0.237$ & $0.261$ & $0.32$ & $<0.32$ &
\multicolumn{1}{l|}{$0.237$} & $0.427$ & $0.447$ & $0.66$ \\
\multicolumn{1}{l|}{bcc} & $0.217$ & $0.221$ & $0.247$ & $0.32$ & $<0.32$ &
\multicolumn{1}{l|}{$0.239$} & $0.414$ & $0.432$ & $0.48$ \\
\multicolumn{1}{l|}{fcc} & $0.210$ & $0.217$ & $0.241$ & $0.62$ & $-$ &
\multicolumn{1}{l|}{$-$} & $0.409$ & $0.427$ & $0.38$ \\ \hline
\end{tabular}
\end{center}
\end{table}

\end{document}